\documentstyle[12pt]{article}
 \begin{document}
\topmargin-1.0cm\textheight=22.7cm
\leftmargin-1.0cm\textwidth=16cm
\hsize=6.0truein
\vsize=10.0truein
\parskip=0.0pt
\parindent=15pt
\baselineskip=18pt
\begin{titlepage}
\begin{center}
{\huge  The Gravitational Vavilov--Cherenkov\\ \vspace{0.5cm} Effect}
\footnote{This essay received an ``honorable mention'' from 
    the Gravity Research Foundation, 1998 - Ed.}  
\end{center}
 
\vspace{5mm}
 
\begin{center}
Alberto Saa\footnote{e-mail: asaa@ime.unicamp.br}\\
and\\
Marcelo Schiffer\footnote{e-mail: schiffer@ime.unicamp.br} \\
Departamento de Matem\'{a}tica Aplicada,\\
IMECC, Universidade Estadual de Campinas, \\
C.P. 6065, 13081-970 Campinas, SP, Brazil

\end{center}
 
\vspace{10 mm}
 
\begin{center}
{\large Abstract}
\end{center}
 
In this essay we show that an uncharged black--hole
 moving  superluminally in a transparent dielectric medium violates Hawking's
area theorem. The violation is overcome through the emission of radiation. 
Since modes
cannot emerge from the black hole itself, this radiation must 
originate from a 
collective  effect in the medium, in complete analogy with the 
Vavilov--Cherenkov effect. However,  
because the black--hole is uncharged, the emission mechanism must be
different.
We discuss the physical origin of the effect and
obtain a Newtonian estimative. Then we obtain the appropriate equations in
the relativistic case and show that the field which is radiated away is a
combination of gravitational and electromagnetic degrees of freedom. Possible
astrophysical relevance  for the detection of primordial black--holes
and binary systems is discussed.
\end{titlepage}
\newpage

\newcommand{\rot}{\vec{\nabla} \times }
\newcommand{\Div}{\vec{\nabla} \cdot }
\newcommand{\grad}{\vec{\nabla}}
\newcommand{\be}{\begin{equation}}
\newcommand{\ee}{\end{equation}}
\newcommand{\der}[2]{\frac{\partial #1}{\partial #2}}
\newcommand{\h}[1]{h_{#1}}
\section*{The Classical Vavilov--Cherenkov effect}
In one of the interesting twists in the development of scientific ideas, 
in 1904 Sommerfeld calculated the energy radiated away by a charged particle 
moving superluminally in the vacuum -- a result that was buried with the
 advent of special relativity a few years later. When Cherenkov and Vavilov
 observed  the radiation emitted by a charged particle moving superluminally
 in a {\it medium} they were unaware of Sommerfeld's calculation and the
 phenomenon found only the  proper theoretical explanation by the hands of
 Tamm and Frank in 1937
\cite{guinzburg,guinzfrank}.

The phenomenon can be easily understood from heuristic considerations.
 Think of a system moving inertially with velocity $\vec{v}$ through a
 medium of refraction index $n$. As $N$ particles are emitted by the
 system , its internal energy changes by the amount:
\be
\Delta M = - N \hbar(\omega - \vec{v} \cdot \vec{k})
\ee
where $\omega$ and $\vec{k}$ are the frequency and wave vector of the emitted 
photons, and back-reaction effects were neglected. 
Because $ \vec{v} \cdot \vec{k} = \omega n v \cos \theta$, 
where $\theta$ is the angle formed between the directions of propagation of 
the photon and that of motion of the particle,
\be
\Delta M = -N \hbar\omega(1- n v \cos \theta)
\ee
Similarly, the absorption of $N'$ photons
is followed by a change in the system's internal energy:
\be
\Delta M' = N' \hbar\omega(1- n v \cos \theta)
\ee
An elementary particle is a structureless system, thus
 $\Delta M=\Delta M'=0$. If the particle moves subluminally then
 $ n v < 1$ and it can neither emit nor absorb a photon from the 
environment. In contrast, if it moves fast enough such that the 
{\it Ginzburg--Frank} condition $n v > 1$ is met, then it can either
 emit or absorb photons provided they lie on the  conical surface
\be
\cos \theta_0 = \frac{1}{n v}
\ee
What is actually  observed is light emitted from the whole interior of the
Cherenkov cone. The reason for this is that the charged particle in a 
dielectric cannot be thought of as structureless as it creates
a polarization cloud that is dragged  alongside its motion. In a process
involving absorption and emission of quanta in a given field mode, the total
entropy change of polarization cloud is
\be
\delta S_{\mbox{\small cloud}}= T^{-1}(N' - N) (1-n v \cos \theta) \, 
\ee
where $T$ is the temperature of the medium, here assumed homogeneous.
 The entropic bookkeeping should also include the entropy of the absorbed
 and emitted photons. For intense radiation the entropy change in the 
radiation behaves as
$\delta S_{\mbox{\small rad}} \approx \log (N'/N)$, which is much 
smaller than 
the entropy change in the polarization cloud and can be 
neglected \cite{bekschi}. 
Thus the second law requires that 
\be 
(N' - N) (1-n v \cos \theta) > 0 \, .
\ee
Accordingly, in the absence of incoming radiation $N'=0$,
the second law says that modes lying  inside the Cherenkov cone  are
 spontaneously emitted. {\it En passant}, we mention a new effect discussed
 recently 
in which there is some incoming radiation. In such a circumstance 
the second law imposes 
$ N > N'$, that is to say, the outgoing radiation is amplified for modes
 lying inside the Cherenkov cone, an effect so far not
 observed \cite{bekschi}. The present discussion makes
 transparent the fact that the light emission mechanism is
 a collective effect of the medium.

On  a more formal basis, the Cherenkov radiation is derived from Maxwell's
equations for a moving particle as a source \cite{landau}:
\begin{eqnarray}
\Div \vec{B}= 0 & &\rot \vec{B} + \frac{1}{c}\der{\vec{E}}{t} = 0 \nonumber \\
\Div \vec{D} = 4 \pi \, Q \, \delta^3(\vec{x}-\vec{v}t) & &
\rot {H} - \frac{1}{c}\der{\vec{D}}{t} = 4 \pi  \vec{v} \, 
Q \, \delta^3(\vec{x}-\vec{v}t)  ,
\label{Maxwell}
\end{eqnarray}
with $ \vec{D} = \hat{\varepsilon} \vec{E}$, where $\hat{\varepsilon}$
 is the dielectric operator (this expression stands for the Fourier
 transform of 
 $\vec{D}(\omega) = \varepsilon (\omega) \vec{E}(\omega)$) and
  $\vec{B} = \vec{H}$. The stopping power $dF$, the energy radiated
 per unit of path length
 per frequency interval that follows from these equations is \cite{landau}
\be
dF = -\frac{i e^2}{\pi} d\omega    \left( \frac{1}{c^2}- 
\frac{1}{\varepsilon v^2}\right) \sum \omega \int \frac{d \xi}{\xi}
\label{stopping}
\ee
where the summation refers to the two branches 
$\omega = \pm \left|\omega\right|$ and
 $\xi \equiv q^2 -\omega^2 (\varepsilon/c^2 -1/v^2)$ with
 $q$ standing for the transverse momentum. Although the medium is transparent
it has always a (very) small imaginary part 
$\Im (\omega)\stackrel{>}{<} 0$ for $\omega \stackrel{>}{<} 0$, a
 consequence of causality. Because 
for $v > c/\varepsilon$
the poles are located beneath/above the real axis for 
$\omega \stackrel{>}{<} 0$  the contour integral is indented 
 such that it is closed in the upper/lower plane
for $\omega \stackrel{>}{<} 0 $. In this case, upon summing both
 contributions the real parts drop out and the final result is \cite{landau}
\be
dF= \frac{e^2}{c^2} \left( 1-\frac{c^2}{n^2 v^2}\right) \omega d\omega
\label{stop}
\ee
Note that for $v < c/\varepsilon $, the integral vanishes rendering crystal
 clear the  relevance of the Frank--Ginzburg condition. 

\section*{Uncharged black--holes moving superluminally}

 The discussion in the previous section on the amplification of
 incoming radiation for modes lying within the Cherenkov cone is 
reminiscent of superradiant amplification of radiation by charged/rotating
 black holes and  brings to our mind the {\it gedanken} experiment in which the
 charged particle is replaced by a neutral black hole moving inertially
 but superluminally through a dielectric but transparent medium with 
refraction index $n(\omega) > 1$.  In order to avoid accretion of matter
 by the hole, imagine a solid  dielectric in which a narrow straight
 channel is drilled through which the hole traverses. What happens? Since
 the black hole is uncharged the sole effect is a tidal wake inside 
the medium  that propagates alongside the black hole. Is that all?
That something novel happens here is a direct consequence of Hawking's area
theorem \cite{hawking}.  The  black hole area change due to the 
absorption of  $N$ photons from the medium in a mode  ($\omega , \vec{k}$) is
\be
\Delta A_{\mbox{bh}} = 32 \pi M \Delta M = 32 \pi \omega \, N \, 
(1- n v \cos \theta) \hbar
\ee
{\it Accordingly, the absorption of photons in modes lying inside the Cherenkov cone leads to a violation of the area theorem! } In vacuum the unique 
scenario which does not violate Hawking's area theorem is the spontaneous
 emission ($N < 0$) of quanta
in Cherenkov modes. We reach to the conclusion that {\it  uncharged 
black--holes moving superluminally in a medium spontaneously emit radiation}.
 This violation was recently discussed in \cite{bekschi} where the
emission mechanism was made transparent.  Since waves cannot
 classically emerge from within the hole we must look for their source in  
a collective effect in the medium, like in Vavilov--Cherenkov effect, but
must differ from that one because the black--hole is uncharged.  Clearly
 in the conversion of kinetic energy to waves, gravitation must play the
 pivotal role.
{\it Because gravity pulls on the positively charged nuclei in the dielectric
 stronger than on the enveloping electrons an electric field develops inside
the atom neutralizing the gravitational pull and preventing ionization. It
 is in this way that  an electrical polarization of the dielectric emerges.
 The polarization cloud that dresses the black--hole and moves alongside
 the black-hole 
should be viewed as the true source of radiation. } This argument makes 
transparent the fact that the black hole character of the moving source
 is immaterial here as what  really counts is its gravitational pull, 
so long  as both are much smaller than the channel's width. An 
 ordinary object with the same mass would have similar effect as a black hole.

We can render this intuitive picture quantitative by noting that in the 
Newtonian limit  the induced polarization $\vec{E}_0$ is the field that
 balances the gravitational pull $\vec{g}$:
\be
 e\vec{E}=-\delta \mu\, \vec{g} \, ,
\ee
 where $\delta \mu\approx A m_p$ is the nucleus--electron mass
difference ($A$ is the mass number of the atoms, $m_p$ the proton's mass),
 and $e>0$ the unit of charge.  From the gravitational Poisson
 equation it follows that 
\be
\Div \vec{E}_0=4\pi G M(\delta \mu/e)\delta(\vec{r}-\vec{r}_0)
\ee
 where $\vec{r}_0$ denotes the momentary black hole position.  The electric
 field accompanying the black hole is thus that of a pointlike charge
$Q\equiv GAMm_p/e$.  Note that we need  $|\vec{v}|$ to be
sufficiently large for the Ginzburg--Frank condition to hold 
and consequently we must go
beyond the Newtonian approximation.

\section*{The Gravitational Vavilov--Cherenkov Effect}

In order to evade from complications stemming from matter accretion, we
 drilled a channel in the dielectric whose width is very much larger than
the  Schwarzschild radius. Under this circumstance we can consider Einstein's
 theory in its linearized form.  We focus now our attention on a single 
atom in the dielectric. Let $\vec{u}$ represent the mean velocity of
 the electrons and the nucleus of such an atom averaged over
a time scale large enough to neglect internal motion within the atom but not 
too large to disregard thermal or oscillatory  motions within the 
dielectric  (phonon waves). 
Clearly $|\vec{u}| << 1$ (hereafter we set $c=1$). 
Thus, up  to the first order in the 
velocity the gravitational pull differential in the atom is:
\be
F^i_{\mbox{grav}} = \delta \mu \frac{d^2 x^i}{dt^2} = - \delta \mu 
\left(\Gamma^{i}_{00} + 2 \Gamma^{i}_{0j} u^j\right)
\ee
As we discussed already,  electromagnetic fields develop inside the atom
 in order to prevent ionization. Because the electromagnetic force 
\be
\vec{F}_{\mbox{elect}} = e \left( \vec{E}_0 + \vec{u} \times \vec{B}_0 \right)
\ee
must balance the gravitational pull, we identify
\begin{eqnarray}
e E^i_0 [\vec{u}] &=& \delta \mu \left( \h{0i,j} - \frac{1}{2} \h{00,i} +
 \h{ij,0} u^j \right) \, ;\\
e \epsilon_{ijk} B^k_0[\vec{u}] u^j &=& \delta \mu \left( \h{i0,j} - 
\h{j0,i}\right) u^j \, ,
\end{eqnarray}
where the linearized version of the Christoffel symbols was used. This
 decomposition is unique and dictated by the fact that  the first line
 is associated to a force that does some work while the second line it 
does not. For non-relativistic motion inside the dielectric, we can omit
the velocity dependent term in $E^i_0$, in which case:
\begin{eqnarray}
E^i_0 &=& \frac{\delta \mu}{e} ( \h{0i,0}-\frac{1}{2} \h{00,i}) \, ;
\nonumber \\
B^i_0 &=& \frac{ 2 \delta \mu}{e} \epsilon_{ijk} \h{0j,k} \, .
\label{pol}
\end{eqnarray}

In the gauge  $\overline{h}^{\mu \nu}_{,\nu}=0$, 
the linearized Einstein's equations read
\be
\Box \overline{h}_{\mu \nu} = -16 \pi T_{\mu \nu} \, ,
\ee
where $T_{\mu \nu}$ is the energy momentum of a source of proper mass
 $M$ and velocity $v^\mu$
\be
T_{\mu \nu} = M v_\mu v_\nu \gamma^{-1} \delta^3 (\vec{x} - \vec{v} t) \, .
\ee
For constant velocity of the source, linearized gravity is equivalent to 
a theory involving  a  single scalar gravitational potential $\psi$ :
\be
\h{\mu \nu} = 4(v_\mu v_\nu + \frac{1}{2} \eta_{\mu \nu}) \psi \, ,
\ee
that satisfies
\be
\Box \psi = - 4 \pi M \gamma^{-1} \delta^3 (\vec{x} - \vec{v} t) .
\ee
In terms of this field, the gauge condition translates into
\be
\der{\psi}{t} + \vec{v} \cdot \grad \psi =0  \, ,
\label{convective}
\ee
where $(\vec{v})^i= v^i/v^0$ and $\gamma = v_0^2$ is the Lorentz factor. In the
three dimensional vector notation the polarization reads:
\begin{eqnarray}
\label{p1}
\vec{E}_0 &=& \frac{4 \delta \mu}{e} \left[ \gamma^2 \vec{v} \dot{\psi} -
\frac{1}{2} (\gamma^2 -\frac{1}{2}) \grad \psi\right] \nonumber \\
\vec{B}_0 &=& \frac{8 \delta \mu \gamma^2}{e} \vec{v} \times \grad \psi
\end{eqnarray} 

Like in pyroelectric media, the electromagnetic 
fields are the sum of the polarization given by eq. (\ref{p1}) 
and the perturbations $(\vec{e},\vec{b}), \vec{E} = \vec{E}_0 +\vec{e}; 
\vec{B}=\vec{B}_0 + \vec{b}$. In terms of the new fields and in the
absence of external charges or currents, the first pair of Maxwell's 
equations read\footnote{In what follows, 
$\frac{1}{\hat{\varepsilon}} f(t) = \frac{1}{\hat{\varepsilon}}
\int \tilde{f}(\omega) e^{-i\omega t} 
d\omega =  \int\frac{ \tilde{f}(\omega)}{\varepsilon(\omega)} 
e^{-i\omega t}
d\omega$}

\begin{eqnarray} 
\Div \vec{d} &=& \frac{4 \hat{\varepsilon} \delta \mu}{e}
\left[ \gamma^2 \ddot{\psi} + \frac{1}{2} (\gamma^2 -\frac{1}{2}) 
\nabla^2 \psi\right] \, ;\\
\rot \vec{b} -  \der{\vec{d}}{t} &=& \frac{4\hat{\varepsilon} \delta \mu}{e}
\left[\gamma^2 \vec{v} \ddot{\psi} - \frac{1}{2} (\gamma^2 -\frac{1}{2}
 \grad \dot{\psi}) -\frac{2 \gamma^2}{\hat{\varepsilon}}
 ( \vec{v} \nabla^2 \psi + \grad \dot{\psi})\right], 
\end{eqnarray}
where $\vec{d} = \hat{\varepsilon} \vec{e}$. Inspired by the Newtonian discussion,
 we wished to interpret the right hand side
of these equations as the effective charge density and current of the 
polarization cloud. The equation $\Div \vec{B}=0 \rightarrow  
\Div \vec{b}=0$, lends support to this strategy. Unfortunately, 
the remaining equation
\be
\rot \vec{e} + \der{\vec{b}}{t} = - \frac{4 \gamma^2 \delta \mu}{e} \,
 \vec{v}\, \times \, \grad \dot{\psi}
\label{offending}
\ee
is not homogeneous preventing such and identification. We overcome this
 hurdle by defining a new pair of fields $(\vec{\epsilon},\vec{\beta})$
 which combine together electromagnetic and gravitational degrees of freedom:
\begin{eqnarray}
\vec{\epsilon} &=& \vec{e} -\frac{4 \gamma^2 \delta \mu}{e} \, \vec{\sigma}\\
\vec{\beta}  &=& \vec{b}  -\frac{4 \gamma^2 \delta \mu}{e} \, \vec{\theta}
\end{eqnarray}
where
\begin{eqnarray}
\vec{\sigma} &=& \left(\frac{1}{2} -\frac{1}{4\gamma^2} +
 \frac{2}{\hat{\varepsilon}}\right) \, \grad \psi + \frac{2-\hat{\varepsilon}}{2} 
\dot{\psi} \, \vec{v}\\
\vec{\theta} &=& 2 \frac{1-\hat{\varepsilon}}{\hat{\varepsilon}} \, \vec{v} \, \times \, 
\grad \psi
\end{eqnarray}

In terms of these new fields the offending term in eq.(\ref{offending}) is
 removed and the field equations attain Maxwell's form
\begin{eqnarray}
\Div \vec{\delta} &=& 4 \pi \, \overline{Q} \, \delta^3(\vec{x} -\vec{v} t) \,
 ;\\
\rot \vec{\beta} - \der{\vec{\delta}}{t} &=& 4 \pi \, \overline{Q} \,
 \vec{v} \delta^3(\vec{x} -\vec{v} t) +\vec{j}\, ;\\
\Div \vec{\beta} &=&0 \, ;\\
\rot \vec{\epsilon} + \der{\vec{\beta}}{t} &=& 0 \, ,
\end{eqnarray}
where  $\vec{\delta} = \hat{\varepsilon} \vec{\epsilon}$,
\be
\overline{Q} = \frac{8 M \delta \mu }{ e^2 M_p^2} \, e \, ,
\ee
with $e$ the elementary charge and 
\be
\vec{j} = - \frac{(1-\hat{\varepsilon}) \, \gamma \, 
\overline{Q}}{ 4 \pi M \epsilon} \left( \vec{v} \, \nabla^2 \psi + 
\grad \dot{\psi}\right) \, ,
\ee
a divergence free current (by virtue
of equation (\ref{convective})).

Our new set of equations differ from eqs.( \ref{Maxwell}) in that they have
 an additional current $\vec{j}$. Nevertheless, this extra term is not
 responsible for additional work on the moving particle  because of 
$\vec{j} \cdot \vec{v} =0$ [see eq.(\ref{convective})] and the linearity
 of the differential equations. Put in other words, this term is not
 responsible for energy dissipation and the formula for the stopping
 power [eq.(\ref{stopping})] applies to these equations as well under
 the replacement\footnote{One could have  wondered whether this 
application of the stopping power formula is legitimate because  a 
proper evaluation of the power dissipated by the source should 
involve both electromagnetic and gravitational `frictions', something
 we did not give its due care. Some reasoning reveals that after all
 our result is correct. Because the above `electromagnetic equations'
 can be derived from a variational principle involving a vector potential
 $\alpha_\mu$ with an interaction
 term $ \overline{Q} \int \alpha_\mu dx^\mu$, the variation of
this term with respect to the path of a test particle path yields a 
Lorentz force 
 $\vec{F}_{\mbox{\tiny friction}} = \overline{Q}(\vec{\epsilon}+
 \vec{v} \times \vec{\beta})$. Accordingly, gravitation and
 electromagnetic `frictions' 
are automatically taken care of, validating our procedure.}
 $Q \rightarrow \overline{Q}$.

Summarizing our results we say that whenever a gravitating object moves
 superluminally in a dielectric medium and the Frank--Ginzburg condition
 is met,
then the composite system (object+medium) will radiate an admixture of
 gravitational and electromagnetic degrees of freedom. This is what
 we call the gravitational Vavilov--Cherenkov effect. It is important
 to remark here that the non-relativistic approach missed a Lorentz 
factor in the effective charge $\overline{Q}$ of the moving source.
 In turn, this factor is responsible to an enhancement of the
 energy radiated away by the very same factor squared [see eq.(\ref{stop})].
Could the effect have practical importance or is it merely of academic
 interest?
Because $Q/e$ is about $10^3A$ times the gravitational radius of the
hole measured in units of the classical radius of the {\it electron\/},
 a fast $10^{15}$ g primordial black hole moving in a suitable dielectric
would radiate just like an equal fast particle bearing  $\sim 10^3 A$
elementary charges.  This is relevant for the experimental search for
primordial black holes. Furthermore the Newtonian approach indicates an 
analog of synchrotron radiation for a gravitating system in circular 
motion. This might be of astrophysical importance in context of binary systems.

As a concluding remark, we emphasize that our argument involves a big 
assumption,  that the dielectric has time to relax to form the above
 compensating field.  Such relaxation does occur for sufficiently small 
$|{\bf v}|$, but since we need $|{\bf v}|$ to be sufficiently large for 
the Ginzburg--Frank condition to hold, stringent conditions are required
 of the dielectric (high $n$ and fast relaxation). Thus, the form of the 
dissipation could differ considerably from the one here discussed for 
media that do not relax fast enough. 

\section*{Acknowledgements:} One of us (M. S.) is deeply indebted to 
J. D. Bekenstein for many enlightening discussions. We are thankful to 
S. Oliveira
for carefully reading the manuscript. This work was supported by
FAPESP and CNPq.

\end{document}